%
%
\input cp-aa.tex
\input psfig

%
\MAINTITLE{Evolution and dynamics of poor clusters of galaxies}
\AUTHOR{Gast\~ao B. Lima Neto\PRESADD{Observatoire de Lyon, France}
        and Frank W. Baier}
\INSTITUTE{Universit\"at Potsdam,
 c/o Astrophysikalisches Institut Potsdam,\newline
 An der Sternwarte 16,
 14882--Potsdam, Germany
}
\DATE{ Received ???? ; accepted ????? } 
\OFFPRINTS{ G. B. Lima Neto }
\ABSTRACT{We have performed $N$-body simulations of poor clusters with enough
particles to resolve the individual galaxies. The main cases of
initial conditions are: (i) 50 equal mass galaxies;
(ii) 60 galaxies with masses drawn
from a Schechter distribution function, and (iii) the collision of two 
subclusters
each one containing 25 galaxies. The evolution and kinematics of the first
ranked galaxy, the substructures, and the possible rotation of 
clusters are investigated.

A massive object with cD characteristics is always formed and is
never found farther than 100 kpc from the centre of the cluster. This
massive galaxy oscillates around the cluster centre with an average
peculiar velocity of 70 km~s$^{-1}$.
We show that increasing the initial intra-cluster medium (ICM)
mass while keeping the galaxies mass
and structure without change, raises the merging rate due to the dynamical
friction with the ICM.
Substructures are almost always present on the galaxy count contour plots.
The
ICM projected density, which should be similar to the X-ray emission map, presents
strong substructures when we have colliding subclusters. Otherwise, in isolated
clusters, the substructure
is less pronounced indicating that substructures should reflect important
but transient dynamical phenomena.
We propose that clusters formed by an off-centre collision and subsequent
merging of two approximate equal mass subclusters should show a general
rotational pattern that
could be detected even after the relaxation of the cluster. These clusters
would have a spin parameter, $\lambda$, of about 0.3--0.5.
}

\THESAURUS{11.03.1, 11.06.1, 11.05.1, 13.25.2, 03.13.4}
\maketitle
\AUTHORRUNNINGHEAD{G. B. Lima Neto \& F. W. Baier}
\titlea{Introduction}

Poor clusters of galaxies are important bound gravitational structures
with velocity dispersion of 300--600 km~s$^{-1}$,
that are intermediate between groups (a few tens of galaxies) and
rich clusters (many hundreds of galaxies). Poor clusters were first
catalogued by Morgan et al. (1975) and Albert et al. (1977), and were
shown to have optical and X-ray properties that are a smooth
continuation of the characteristics of rich clusters of galaxies
(Bahcall 1980). Recent studies (e.g. Dell'Antonio et al. 1995) have
shown that the galaxy distribution in poor clusters reflects the X-ray
distribution. Moreover, the mass in galaxies is about 5--10 \% of the
total mass and about half the gas mass.

As aggregates of galaxies, poor clusters provide
some ideal conditions for the study of galaxies in `community'. Given their
velocity dispersion it is natural to expect that galactic interactions play a 
influential
r\^ole in the galaxy and cluster evolution. Indeed, many poor clusters
have a cD galaxy near its centre, that can be defined either by the galaxy
distribution
or the X-ray emission. There are two main theories for the formation
of cD galaxies in clusters. First, they may be the result of a cooling flow
of the intra-cluster gas at a rate of some tens of solar masses per year
that piles up at the bottom of the cluster
potential wells (e.g. Fabien et al. 1984). The second possibility is the
formation by galactic cannibalism (Ostriker \& Tremaine 1975) which
is expected when phenomena like dynamical friction, tidal stripping, and
mergers are common. Galactic cannibalism is often supported by the observation 
that at least half the observed cD galaxies have more than one nucleus, and the
fact that the formation of central
giant galaxies is easily obtained in $N$-body simulations of groups and 
clusters (e.g. Barnes 1988, Bode et al. 1994). Moreover, the extended envelope
around cD galaxies can be explained by the accumulation of tidally stripped matter
from galaxies at the bottom of the cluster potential wells (Merritt 1983).

Thus, although poor clusters in itself are important structures that
bridges the well studied rich clusters and groups of galaxies, they may
be also important if they are considered as the building blocks of
rich clusters. Indeed, it is well known that many rich clusters have
clear signs of substructures. The frequency and intensity of this
subclustering, however, are a subject of debate. For instance,
Baier et al. (1996) advocate that almost all clusters show clear
substructures in the galaxy and X-ray distributions. On the other hand,
Jones \& Forman (1990) estimate that the fraction of clusters with double or
multiple maxima in the X-ray distribution is 30\%, Geller \& Beers (1982)
found about 40\% of clusters having multiple peaks in the galaxy distribution,
and the analysis of gravitational lens can suggest that the
total mass distribution (i.e. including the invisible matter) may have
substructures (Fort \& Mellier 1994).

The systematic presence of multiple peaks in the 
distribution of the various components of clusters may be understood if clusters 
are formed by the merger of smaller units, that is, poor clusters of galaxies
(McGlynn \& Fabian 1984). Zabluboff and Zaritsky (1995) presented X-ray and
optical evidence that the two substructures (separated by about 0.7 Mpc in
the plane of the sky, with line-of-sight relative velocity of about
100~km~s$^{-1}$) in the cluster Abell 754 are in the process of colliding.
The cluster Abell 569 seems also to be formed by two subclusters that may
be falling in a spiral towards each other (Beers et al. 1991,
Baier et al. 1996). Finally, Ulmer et al. (1992) show that the centers of the
X-ray and galaxies distributions of Abell 168 are probably disjoint.
They
interpret this as an evidence that this cluster was formed by
a collision of two equal sized subclusters.

It is not clear however how these
mergers of poor clusters may affect the intra-cluster medium (X-ray
emitting gas and dark matter), the 
galaxies, and their
relationship. Besides the dynamical process that operates in isolated virialized 
clusters like dynamical friction, two-body relaxation, tidal effects, and 
mergers of galaxies, there are also the effects due to encounters of clusters. 
Namely, we have the shock and eventual heating of the X-ray emitting gas, and 
the 
temporal variation of the total gravitational potential, leading to a violent 
relaxation of all components of the cluster. Another important point to consider 
is the possibility that the collisions of subclusters may not always be head-on
but parabolic. In this case, and if the subclusters merge within a Hubble time,
the final cluster may present a global angular momentum that may be determined 
observing the radial velocity of the galaxies.

Although theoretical models can describe fairly well and uncover the basic 
physics of `well behaved' clusters -- i.e. objects close to virial equilibrium, 
without 
strong substructure -- (e.g. Merritt 1983), one needs numerical methods to 
follow 
in a self-consistent way the non-linear evolution of galaxies in clusters.

The present study addresses the dynamical evolution of isolated poor clusters 
as well as the collision of two clusters.
In this work we will concentrate on the evolution of the first ranked
galaxy of poor clusters, its formation and kinematics near the centre of the 
cluster; the
evolution of the Mass (Luminosity) Function of the galaxies; and the
presence of eventual substructures that may appear during the evolution.

Our main tool in this research is the use of self-consistent $N$-body 
simulations. This
technique enables us to follow the time evolution from a given set of
initial conditions to the present configuration. Our aim is to investigate
different phases of the evolution of poor clusters of galaxies, including the 
encounters of subclusters, searching
evidence for interactions among galaxies that occurred in the past and 
that may be responsible for the features observed today.

This paper is organized as follows: in section 2 we describe the techniques 
employed and the initial conditions that defines our models. In section 3 we 
describe our results concerning the formation of the first ranked galaxy (FRG),
the presence of 
substructures (in the galaxy and the total mass distribution). In section 4
we discuss our results in connection to previous theoretical and observational 
studies and we conclude in section 5.

\titlea{Method}

\titleb{Technique}

We have used $N$-body simulations in order to study the evolution of
galaxies in clusters. The equations on motion are integrated with the
Tree-Code developed by Barnes \& Hut (1986) and ported to {\sc fortran} 
and vectorized by
Hernquist (1988). The Tree-Code is particularly well adapted for
the simulations of granular systems (such as a cluster of galaxies),
without imposing any geometric symmetry since it is of lagrangian type
(i.e. gridless). We have used a time step of 0.25 time-units,
\fonote{The units used to express the results
of the simulations are scaled as ($G = 1$): 
[length] = 1.0~kpc, [time] = $6.325\times10^6$~yrs, [mass] 
= $5.56\times10^9~M_\odot$, [velocity] = 154.6~km/s, 
[$\mu$] = $5.56\times10^3~M_\odot/{\rm pc}^2$ and [$\rho$] = 
$5.56~M_\odot/{\rm pc}^3$.}
 tolerance parameter of 0.75, and quadrupole correction. The softening
parameter of each particle is equal to 0.5
length-units (or kpc, with our adopted scaling), a compromise value between
obtaining runs with good energy 
conservation and the resolution need to well resolve the core of galaxies. With 
these parameters,  the energy conservation is about 1\% for
8400 time steps (corresponding to $13.3\times10^9$ years) using 46\,000 equal
mass particles.

In all runs the masses of the individual particles are the same in order
to avoid spurious mass segregation (heavier particles falling towards
the centre) due to two body relaxation.

\titleb{Initial conditions}

In order to follow the evolution of galaxies in clusters, we have used three
main families of initial conditions in our simulations:
\medskip
\item{a)} Isolated clusters with equal mass galaxies, without ICM.
\item{b)} Isolated clusters with galaxies following a Schechter (1976) luminosity
function, with ICM.
\item{c)} Collision of two equal mass clusters of galaxies, with ICM.
\medskip

Notice that ICM here means a dark intra-cluster medium of {\it collisionless
particles}, not the intra-cluster dissipative X-ray emitting gas.
When it applies (families b and c), an ICM is superimposed to the initial galaxy
distribution. The function of the ICM is to mimic an assumed invisible matter
component.

The initial conditions of our simulations
should be regarded as the epoch between the relaxation of the cluster
just after detaching from the Hubble flow and before appreciable interactions
have affect the galaxies or the cluster itself. In all cases, the simulations are
started with the clusters already
relaxed, close to a
virial quasi-equilibrium state.

The distribution of galaxies in rich clusters is
usually well fitted by the isothermal King profile, which is often
approximated by the analytic King profile
(eg. Sarazin 1988, and references therein). Since poor clusters
seems to be a physical continuation of rich clusters of galaxies
(Bahcall 1980), one would expect that the
King profile would also apply for poor clusters. However, due to
small number statistics, in practice it is difficult to determine
precisely
the galaxy distribution in poor clusters. In any case, the 
analytic King profile is just an approximation of the true King
distribution, the main difference between them being
the asymptotic behaviour. The true King distribution has
a finite mass and extent, while the analytical one is infinite.

For this reason we chose to use a Plummer distribution (which is
fully analytical) in order to
model the initial conditions of our simulated clusters.
The Plummer profile has
a finite mass and is very close to the analytical King profile
for $r \la 10 r_{\rm c}$ (taking into account that the core radius,
$r_{\rm c}$, of
a Plummer sphere should be about 1.75 times greater than the corresponding King
core radius).

Below, we give the details of the
construction of each family of initial condition.

\titlec{Clusters with equal mass galaxies}

In this case, each member of the cluster is a clone of a galaxy that is
modelled according to the following `recipe'. The particles that make up the
galaxy are placed randomly in a Plummer sphere of total mass $M_{\rm tot}$,
$$ \rho(r) = {3 \over 4\pi} {M_{\rm tot} \over r_{\rm c}^3 }
\left[ 1 + \left(r\over r_{\rm c}\right)^2 \right]^{-5/2}\, ,\eqno\autnum$$
having a very low velocity
dispersion, so that the total potential energy is much greater
than the total kinetic energy.
Table~1 summarizes its properties.

\begtabfull 
\tabcap{1}{Initial conditions of the Plummer galaxy used as a clone. The
dynamical time scale, $t_{\rm dyn}$ is defined as $GM^{5/2}/2|E_{\rm
tot}|^{3/2}$ and $r_{\rm c}$ is the core radius }
\halign{\strut\vrule\quad#\hfill\quad & \hfill # &\vrule\quad #\hfill\quad &
\hfill #\quad\vrule \cr
\noalign{\medskip\hrule}
 No points {}$^{\strut }$  & 500   & $M_{\rm tot}$              & 90     \cr
 $r_{\rm c}$               & 20    & Total Energy/$M_{\rm tot}$  & $-$1.27\cr
 $t_{\rm dyn}$ & 31.6{}$_{\strut}$ & $2 E_{\rm cin}/|E_{\rm pot}|$ & 0.095  \cr
\noalign{\hrule\medskip} 
}
\endtab

After generating the Plummer sphere, we followed its evolution in isolation
with the Tree-Code for 800 time-units, about 25 $t_{\rm dyn}$.
Since the object is initially cold ($2E_{\rm cin}/|E_{\rm pot}| <\!\!< 1$) it
collapses on a time scale of $\sim 1/\sqrt{G\,\overline{\rho}}$, about
the order of its dynamical time scale. After the collapse, the sphere
oscillates a few times, and relaxes to a quasi-equilibrium virial configuration.
This dissipationless collapse results thus in a relaxed object structurally
similar to an elliptical galaxy. The half-mass radius decreases to 13.3, about
half the initial value 26.1 ($1.305\times r_{\rm c}$, for a Plummer sphere).
The projected density profile is well fitted by a de Vaucouleurs law.

The clusters of equal mass galaxies are then created by placing clones of
this `elliptical galaxy' on another Plummer sphere, now representing
the cluster. Now,
however, the velocity dispersion of the particles (which represent the
galaxies in the cluster) is
chosen so as that the cluster is in virial equilibrium. The velocity of
each galaxy is drawn randomly from a isotropic maxwellian distribution. Notice
that initially all mass is in the galaxies.

We have done 2 simulations each one with 50 identical galaxies. Both
simulations use the same set of structural parameters (Table~2) except that
the
seed of the random number generator was changed thus changing the actual
distribution in the phase space (positions and velocities of the galaxies).
Random fluctuations
explain the differences on quantities like the total energy and the
velocity dispersion between the two clusters.

\begtabfull 
\tabcap{2}{Properties of the clusters having equal mass galaxies.
$r_{\rm coup}$ is
the initial radius of the cluster, $t_{\rm dyn}$ is the dynamical
time scale, $\ell$ is the initial mean harmonic separation of
the galaxies, and $\sigma$ is the initial velocity dispersion. }
\halign{#\hfill&\quad #\hfill&
                \quad #\hfill&\quad #\hfill&\quad #\hfill \cr
\noalign{\medskip\hrule\medskip}
Run Id & $r_{\rm coup}$ & $t_{\rm dyn}$ & $\ell$ 
&
$\sigma$ \cr
\noalign{\smallskip\hrule\medskip}
AM1    &   1500    &  304  &  571 & 1.88 \cr
AM2    & {\ \tt "} &  443  &  727 & 1.64 \cr
\noalign{\medskip\hrule\smallskip} 
}
\noindent For each clusters the number of particles is 25\,000, the total mass
is 4500, and the core radius is 450.
\endtab

The snapshots of the initial conditions and evolution of the cluster AM1 is
shown as an example on Fig.~1.

\begfigwid 0.0 cm
\psfig{figure=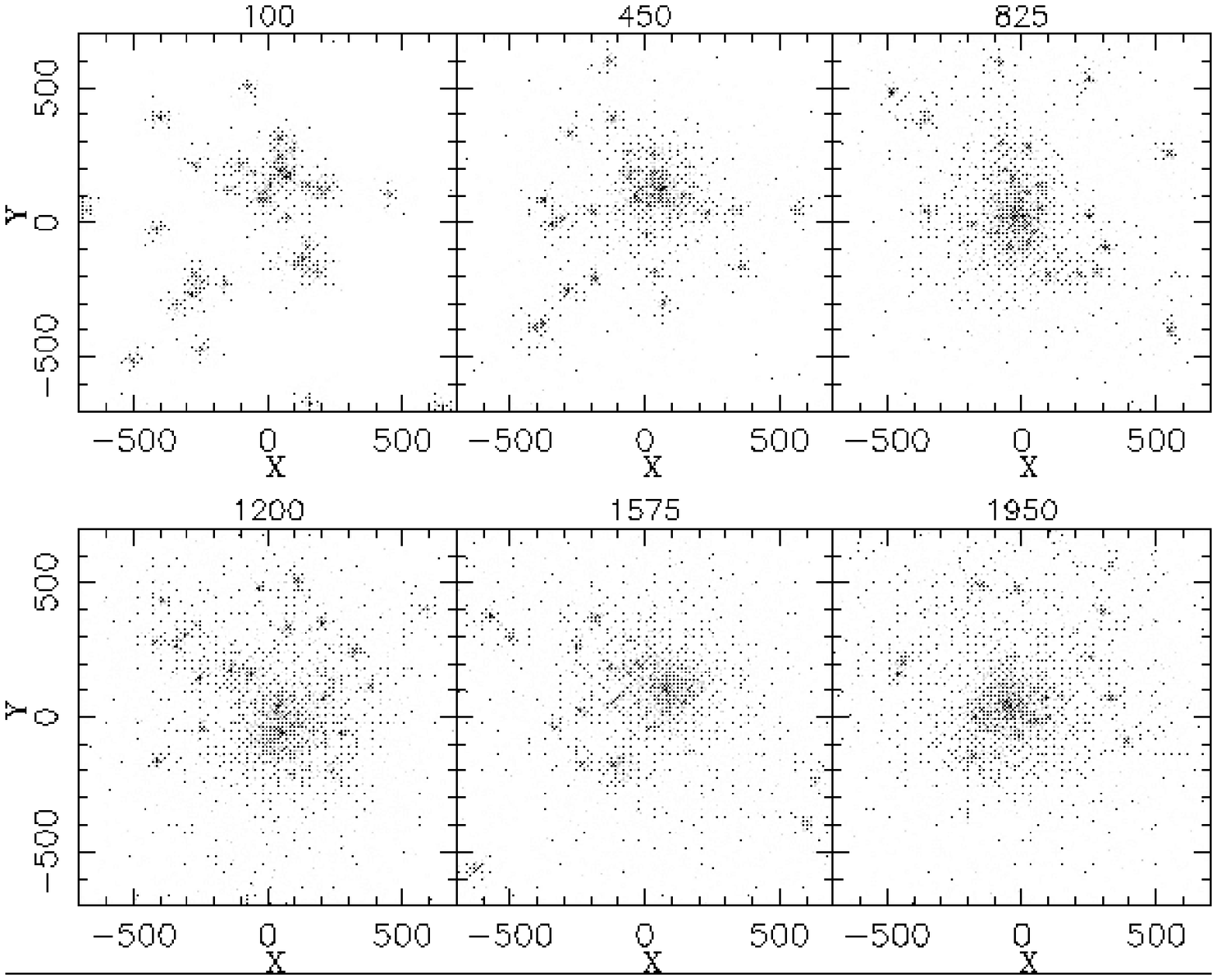,width=18cm}
\figure{1}{
The snapshots of the initial conditions and the evolution of the
simulation AM1 (initially equal mass galaxies and no dark ICM). On top of each
panel is the time. Note that an inter-cluster medium is created due to the
stripping of mass from the galaxies, thus forming an envelope around the
FRG
}
\endfig

\titlec{Clusters with mass (luminosity) function}

Contrary to the previous case, here the galaxies are generated
already in a virial quasi-equilibrium state. The first step however
is determining the masses of the individual objects before actually
generating them.
The masses of the galaxies were randomly drawn from a
Schechter luminosity function
$$ \phi(M) = {\phi_* \over M_*} \left({M \over M_*}\right)^\alpha
\exp(-M/M_*) ,\eqno\autnum$$
with the following parameters: $M_{*}=40$, $\alpha=-1.1$.
We also defined the mass range of the cluster as $M_{\rm min} = 8$ and
$M_{\rm max}=100$. Notice that this range implies a magnitude difference
(assuming constant mass to luminosity ratio) of about 2.7 between
the largest and smallest galaxies.

Having the masses, we generate virialized Plummer spheres scaling each
object with
$r_{\rm core} \propto M^{1/3}$ and making sure that all particles in
the galaxy are gravitationally bound.

Like in the previous case, 60 galaxies were placed randomly in a Plummer
sphere (now the cluster). A dark collisionless ICM is superimposed on the
cluster. The ICM follows also a Plummer density profile and has the same
core radius as the galaxy distribution. The fraction of mass in the ICM
is reported
in Table 3. The masses of the ICM particles are the same of the galaxies
particles in order to avoid spurious segregation.
The galaxy and ICM velocity 
dispersion were so that the clusters were initially in virial equilibrium.
Since the ICM particles are initially in equilibrium with the cluster gravitational
wells, they are not bound to the galaxies individually.
Table 3 describes the initial conditions that we have used.

\begfigwid 0.0 cm
\psfig{figure=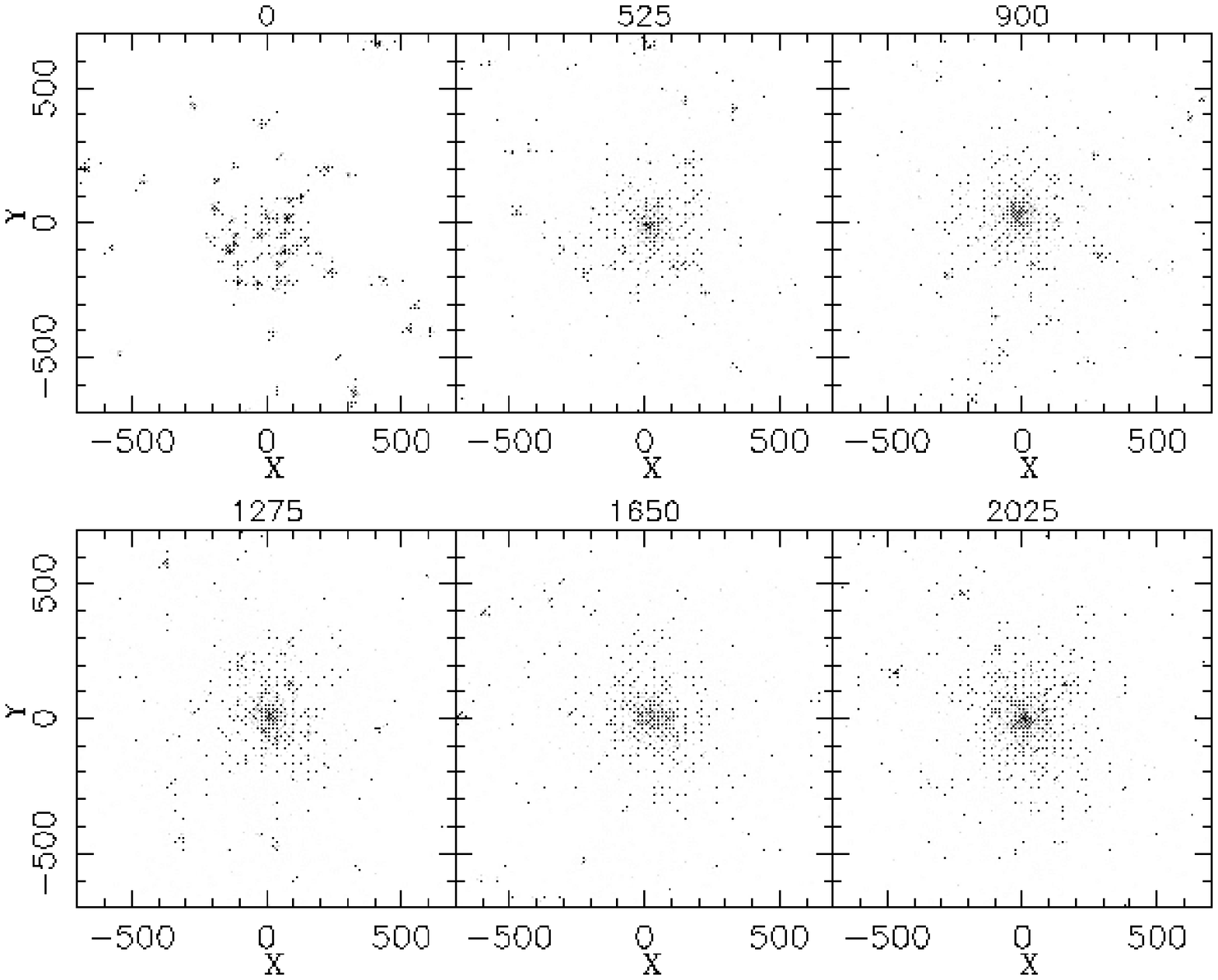,width=18cm}
\figure{2}{
The snapshots of the initial conditions and the evolution of the
simulation AM4 (initially galaxies following a Schechter mass distribution and
0.67\% of the total mass in the form of ICM). On top of each panel is the time.
For clarity, the dark ICM particles are not plotted (but notice that a diffuse
extra galactic component is formed due to the stripping of particles from
the galaxies)
}
\endfig

\begtabfull 
\tabcap{3}{Properties of the cluster having galaxies that follow a mass
function. ICM is the ratio of the ICM mass to the total mass of the cluster, 
$r_{\rm coup}$ is the initial cut-off radius of the cluster, $t_{\rm dyn}$ is
the dynamical time scale of the galaxies,  $\ell$ is the mean harmonic
radius, and $\sigma$ is the velocity dispersion. }
\halign{#\hfill&\quad #\hfill&\quad #\hfill&\quad #\hfill&
                \quad #\hfill&\quad #\hfill&\quad #\hfill\cr
\noalign{\smallskip\hrule\medskip}
Run Id & Mass & ICM & $r_{\rm coup}$ 
       & $t_{\rm dyn}$ & $\ell$ & $\sigma$ \cr
\noalign{\smallskip\hrule\medskip}
AM3  & 2208 &  0.40 &   1200  & 433 & 249 & 1.27 \cr
AM4  & 4423 &  0.67 &{\ \tt "}& 320 & 270 & 1.44 \cr
AM5  & 5519 &  0.67 &{\ \tt "}& 400 & 252 & 1.35 \cr
\noalign{\medskip\hrule\smallskip} 
}
\noindent For each run the number of particles is 45\,000 and the
initial core radius (Plummer model) is 150.
\endtab

The snapshots of the initial conditions and evolution of the cluster AM4 is
shown as an example in Fig.~2, where only the visible particles are shown.

\titlec{Collision of equal mass clusters of galaxies}

\begfigwid 0.0 cm
\psfig{figure=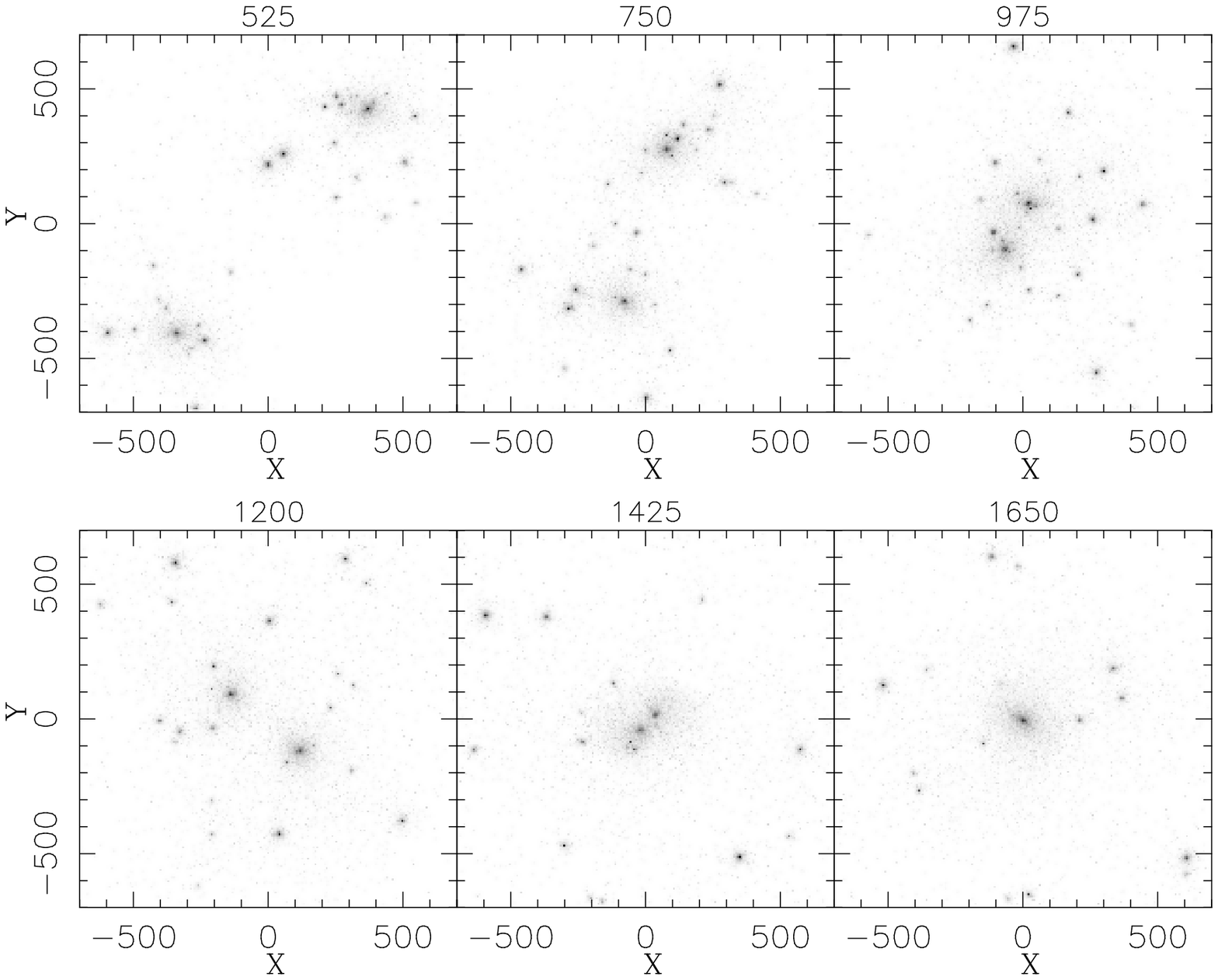,width=18cm}
\figure{3}{
The snapshots of the evolution of the
simulation MS3 (collision of two clusters of 25 galaxies each with the
galaxies following a Schechter mass distribution and
0.67\% of the total mass in the form of ICM). On top of each panel is the
time. As in Fig.~2, the dark ICM particles are not plotted
}
\endfig

For this kind of simulation, we have put two subclusters (or substructures) 
in a collision trajectory. Each subcluster was created already in virial
equilibrium as explained in the above paragraph.
In this case, each subcluster contains 25 galaxies with masses sampled from a
Schechter Function. The properties of the subcluster are given in Tab.~4.

The subclusters are put on a collision trajectory. We have
simulated a head-on and three tangential collisions where the subclusters were
in an initial elliptical orbit.
A summary of the initial collision parameters
is given in Tab.~5. In runs MS1, MS2 and MS3 we use the subclusters SA1 and SA2
with a total of 46\,000 particles. In the MS4 simulation we use the pair
SB1 and SB2 with a total of 70\,000 particles.

\begtabfull
\tabcap{4}{Properties of the subclusters. $M_{*}$ and $\alpha_{\rm Sch}$
are the parameters of the Schechter distribution. $M_{\rm gal}$ and
$M_{\rm tot}$ are the masses in the galaxies and in the whole system,
respectively. $t_{\rm dyn}$ is the dynamical time scale, $\sigma$  is
the velocity dispersion, and $\ell$ is the mean harmonic separation of
the galaxies }
\halign{#\hfill &\quad\hfill # &\quad\hfill # &\quad\hfill #&\quad\hfill #\cr
\noalign{\smallskip\hrule\medskip}
 Models: & SA1 & SA2 & SB1 & SB2 \cr
\noalign{\medskip\hrule\medskip}
Npts & 23\,000& 23\,000& 35\,000& 35\,000\cr
$M_{\rm gal}$ & 623& 602& 602& 623\cr
$M_{\rm tot}$ & 1869 & 1805 & 1807 & 1871\cr
$t_{\rm dyn}$ & 450& 400& 300& 550\cr
$\sigma$ & 0.91 & 0.93 & 1.05& 0.81\cr
$\ell$ & 301 & 241 & 218& 224\cr
\noalign{\medskip\hrule\smallskip} 
}
\noindent For the above models it we set $\alpha_{\rm Sch}=-1.1$, $M_*=40.0$,
ICM=67\%, $r_{\rm core}= 100.0$.
\endtab

\begtabfull
\tabcap{5}{Properties of the systems of two colliding substructures.
`Sep.' is the initial separation between the centre of both substructures,
$v_{\rm rel}$ is their relative velocity, $\epsilon$ and $T_{\rm orb}$ are
the ellipticity and orbital period (oscillation for MS1) of the initial
keplerian orbit. $E/M$ and
$L/M$ are the total energy and angular momentum per mass unit}
\halign{#\hfill&\quad \hfill#&\quad \hfill#&\quad \hfill#&\quad \hfill#
&\quad \hfill#&\quad \hfill#\cr
\noalign{\smallskip\hrule\medskip}
Run Id & Sep. & $v_{\rm rel}$ & $\epsilon$ & $T_{\rm orb}$ & $E/M$ & $L/M$\cr
\noalign{\medskip\hrule\medskip}
MS1& 2500 & 1.5 & 1.00 & 40400& $-0.086$& 0.0\cr
MS2& 2500 & 1.5 & 0.86 & 40400& $-0.086$& 562.5\cr
MS3& 1836 & 1.4 & 0.87 & 7910 & $-0.255$& 315.0\cr
MS4& 1836 & 1.4 & 0.87 & 7895 & $-0.256$& 315.0\cr
\noalign{\medskip\hrule\smallskip}
}
\endtab

The MS1 simulation is a head-on collision of two clusters while
MS2, MS3 and MS4 are tangential collisions. The snapshots of the initial
conditions
and the evolution of the simulation MS3 is shown as an example in Fig.~3.

\titleb{Counting and weighting the galaxies}

Once we run a simulation we must keep track of the galaxies. This is not
trivial for two reasons: the galaxies merge and they have their mass partially
stripped by tidal encounters.

In order to determine in our simulations which particles belong
to a given galaxy, and thus identify each galaxy at a given moment,
we have used a
percolation technique, the so-called `friends-of-friends' algorithm (ex. 
Chincarini et al. 1988). This method is well adapted
to localize the galaxies since it is independent of their shape or
position inside the cluster. Another advantage of this method is that
it selects objects having approximately the same overdensity at
the border compared to the global mean density (West et al. 1988).

Briefly, in the standard percolation algorithm we define a sphere of radius
$r_{\rm p}$ around each particle. The groups of particles with
intersecting spheres are then identified as a `candidate galaxy' in
our simulations. From the group of particles making this
`candidate galaxy', we rejected those that have a velocity
four times greater than the velocity dispersion of the group. In
this way, we eliminated possible unbound particles. Finally,
in order to avoid small groups of particles being identified as a
galaxy, we have only accepted as a galaxy objects with a number of particles
above a given cut-off. This number depends slightly on the simulation
being between 45 to 60 particles (we take into account only the
visible particles, not the particles that forms the dark ICM).

There in no straightforward way to determine a priori the value of
$r_{\rm p}$. Using a too small value we identify any concentration
of particles as a galaxy; with a $r_{\rm p}$ too large, we cannot resolve
galaxies that are close together.
We tried different percolation
radii and retained $r_{\rm p}=4.5$ for all runs.
The objects thus found are in fact the cores of the galaxies since
otherwise it would not be possible to distinguish between close
galaxies.

\titlea{Results}

\titleb{Merging and the first ranked galaxy}

In all simulations we observe the formation of a central giant galaxy by
merging, which we will refer as the first ranked galaxy (FRG).

For the 2 simulations AM1 and AM2 (no ICM, starting with equal mass galaxies)
the merging of galaxies starts at about one $t_{\rm dyn}$ and proceeds steadily
until the end of the simulation (Fig.~4.a). In the AM1 run, the FRG forms
near the centre at about $2t_{\rm dyn}$, and it is well distinct from the
remaining galaxies at $2.5t_{\rm dyn}$. This object will remain at or very
near the centre of the cluster and will slowly cannibalize the other galaxies.
We can still count 29 galaxies at the end of the simulation ($t \approx
12.8\times 10^9$ years). The AM2 simulation, which is intrinsically similar to 
the
AM1 (cf. \S 2.2.1) does {\it not\/} form a giant galaxy near the centre as
quickly as in simulation AM1. An important FRG will only form after 
over than $3t_{\rm dyn}$ and it will only be in the centre of the cluster at
about $4t_{\rm dyn}$ (see also Sec. 3.2).

\begfig 0.0 cm
\psfig{figure=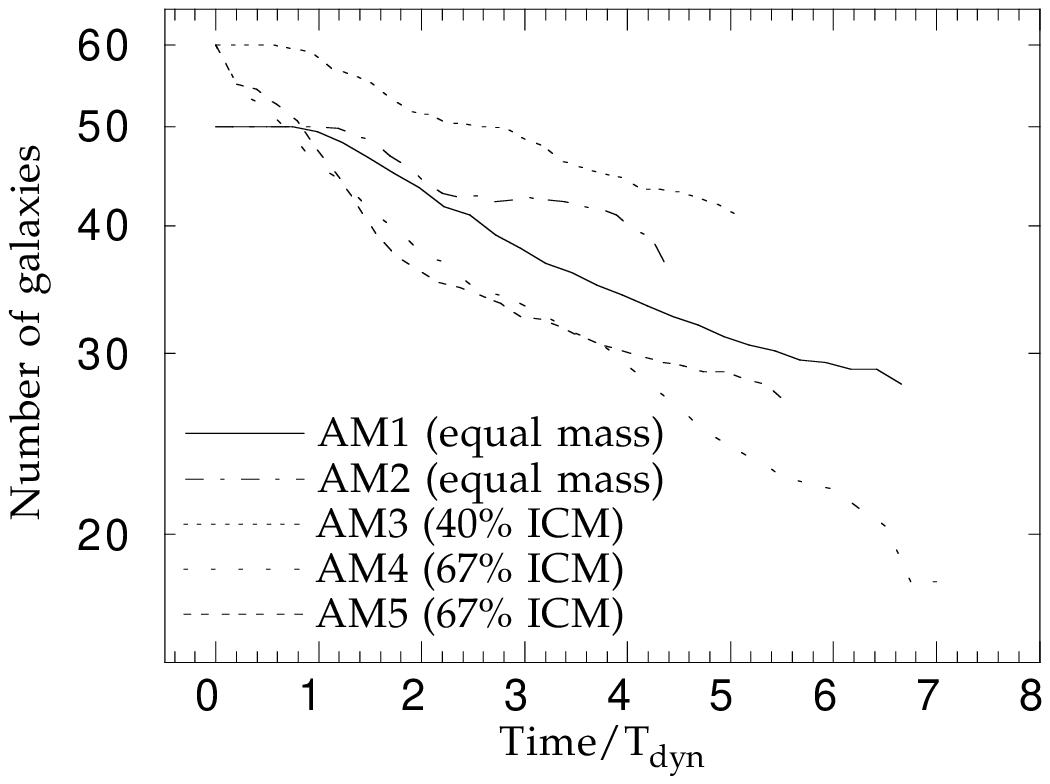,width=8.7cm}
\figure{4.a}{
Evolution of the number of galaxies (merging rate) for the simulations of
isolated clusters. In order to compare, the time scale is normalized by the
dynamical time
scale (see Tabs. 2 and 3). The number of galaxies is plotted in logarithm
scale to emphasize the exponential character.
Note that in some places
curves are not strictly decreasing because when 2 or more galaxies
are too close together, they are counted as one by the `friends-of-friends'
algorithm. Later, when the galaxies separate, they are counted individually 
again
}
\endfig

\begfig 0.0 cm
\psfig{figure=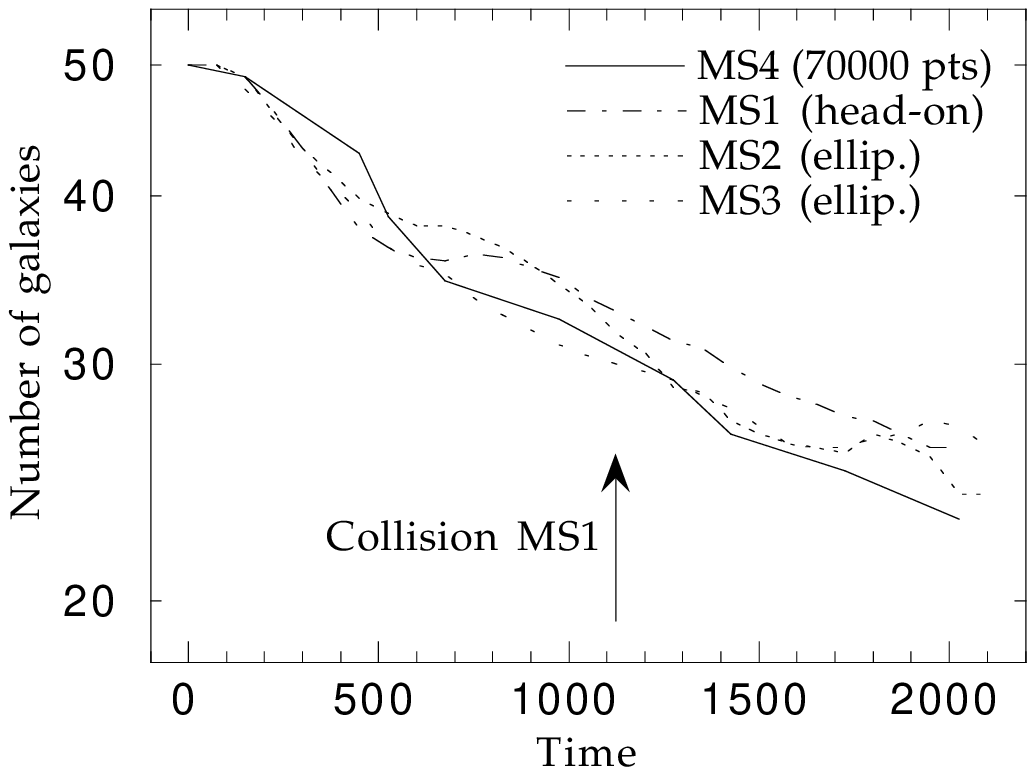,width=8.7cm}
\figure{4.b}{The same as Fig. 4.a for the simulations of collisions of
subclusters. As a reference, it is showed the first collision in simulation
MS1. The initial mean dynamical time scale of the subclusters is 400 time units
}
\endfig

The case of the three simulations AM3, AM4 and AM5 (with ICM, and the masses of
the galaxies following the same Schechter distribution function) is different
from the precedent. Here, there are already a few galaxies which
are naturally (thanks to the mass distribution function) more massive than the
average galaxy
at the beginning of the run. These more massive galaxies fall faster than the 
others
to the centre of the cluster, where they start to cannibalize the smaller
ones. Thus, a central dominant galaxy quickly forms on these
three simulations,
independently on the amount of the ICM. However, the merging rate is
strongly
superior in the more massive cluster (more massive ICM, the mass contained
in the galaxies is about the same for both simulations).
 
For the simulations of merging of substructures, (MS1, MS2, MS3, and MS4)
the formation of
a dominant galaxy is not different from the simulations AM3 and AM4.
Indeed, the merging rate does not seem to change even when the
substructures collide. Each substructure develops its own dominant
galaxy at its centre. These FRGs, however, will promptly merge
when the substructures collide. The time scale of merging of these galaxies is
comparable to the time scale of merging of the ICM of both substructures.

The decrease of the number of galaxies in all the simulations presented here
(Figs. 4.a and b)
are best represented as an exponential decrease, rather than a linear one.
In order to have a quantitative measure, we have combined the number of
galaxies as a function of time for the four MS simulations (collision of two
subclusters) and made a $\chi^2$-fit to a linear and exponential curve 
(Tab.~6).

\begtabfull
\tabcap{6}{$\chi^2$-fit of the combined merging rate of the MS simulations
(collision of subclusters).
`Prob($>\chi^2$)', the confidence of the fit, gives the probability of
having a $\chi^2$ higher than the one
we got. The column
`Fit' gives the equations used to fit the merging rate; $n_0$
is the normalization constant and $b$ is the `inclination' slope in time
units. $t$ is
the time}
\halign{#\hfill\quad &\hfill #\quad &\hfill #\quad &\hfill #\quad\cr
\noalign{\smallskip\hrule\medskip}
Fit & $b$ & $\chi^2$ & Prob($>\chi^2$) \cr
\noalign{\medskip\hrule\medskip}
$n_0 - t/b$ & 84.8 & 76.7 & 87.4\% \cr
$n_0 \exp(-t/b)$ & 2750 & 143.3 & 1.0\% \cr
\noalign{\hrule\medskip} 
}
\endtab

These fits suggest that the merging rate can be expressed as
$\diff N_{\rm gal}/\diff t \propto -N_{\rm gal}$, implying that the number
of mergers in a given cluster is proportional to the number of
galaxies, being higher in the past than now.

\titleb{Mass evolution of the FRG and its envelope}

As the galaxies evolve in the cluster, a fraction of their
mass is stripped by tidal encounters and forms a diffuse
component in the cluster. This diffuse component ends on a huge
envelope around the central dominant galaxy (cf. Figs. 1, 2, and 3).
Here, we analyse the mass evolution of
this stripped matter as well as the mass of the FRG.

Figure 5 shows the mass growth rate of the FRG and the
diffuse component stripped from the galaxies. (The dark ICM particles
are not used to compute the mass in the FRG.)
Both the envelope and the FRG have a strong increase in mass during
the first 1--2 $t_{\rm dyn}$.
The strong evolution of the FRG mass and its envelope
corresponds to the higher merging rate observed at the beginning of the
simulations.
The FRG then basically stops
growing but the envelope goes on accreting stripped matter from
the galaxies.

The reason why the FRG mass growth almost halts is related to
our definition of FRG and envelope. Indeed, what we call the FRG
is the central part of the galaxy (cf. \S 2.3). Therefore, the
FRG mass cannot grow indefinitely. On the other hand, the envelope
is formed by the stripped particles from the all the galaxies. Its
increase reflects the tidal interactions and mergers in the cluster.

\begfig 0.cm
\psfig{figure=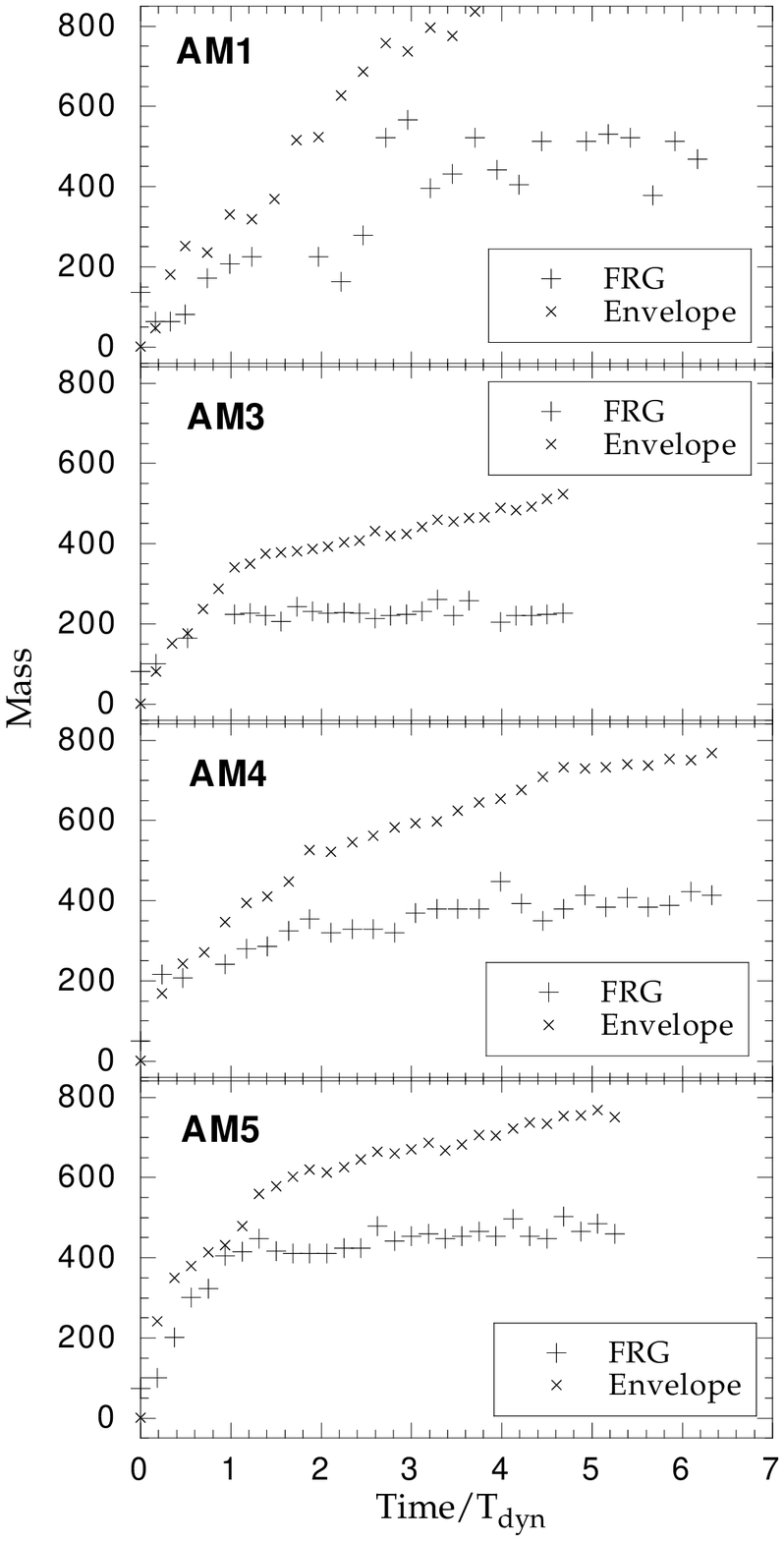,width=8.7cm,clip=true}
\figure{5}{Mass growth rate of the FRG and its envelope.
For comparison purpose, the time is scaled to the
dynamical time scale.
The run ids. are on the top left of each panel}
\endfig

Inspection of Figs. 1, 2, and 3 suggests that the mergers
occur mainly with the central dominant galaxy. In order to
verify this we plot the combined mass of the FRG (the galaxy plus
its envelope) vs.
the number of galaxies in the cluster (Fig. 6). Indeed, this plot
shows a strong correlation between the growth of the total
mass of the FRG and the rate of merging, thus strengthening our
suspicion that mergers are mostly related with the central
dominant galaxy.

\begfig 0.0cm
\psfig{figure=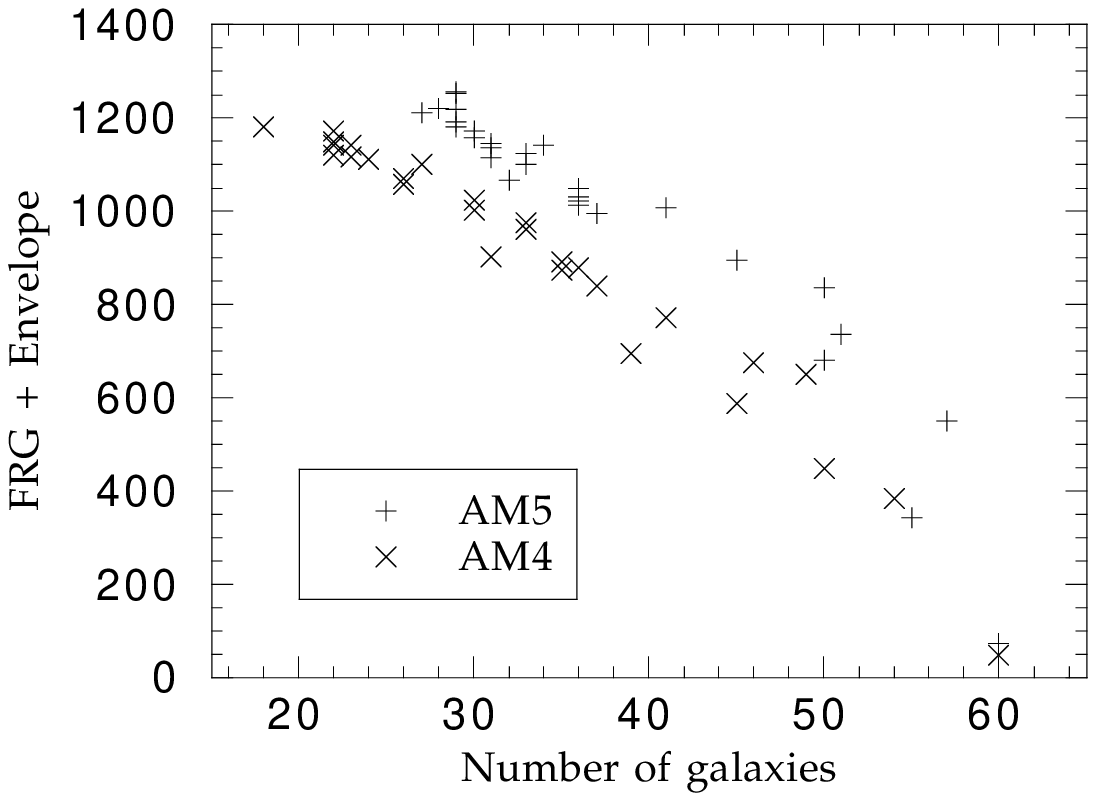,width=8.7cm}
\figure{6}{Total mass of the FRG (galaxy + envelope) as a function
of the number of galaxies for two simulations, AM4 and AM5}
\endfig

\titleb{Position and velocity of the FRG}

As it was seen above, the FRG is always near the centre of the cluster. In
Figs.~7.a and 7.b we show the position of the FRG as a function of time for the
simulations of isolated clusters and colliding clusters.

\begfig 0.cm
\psfig{figure=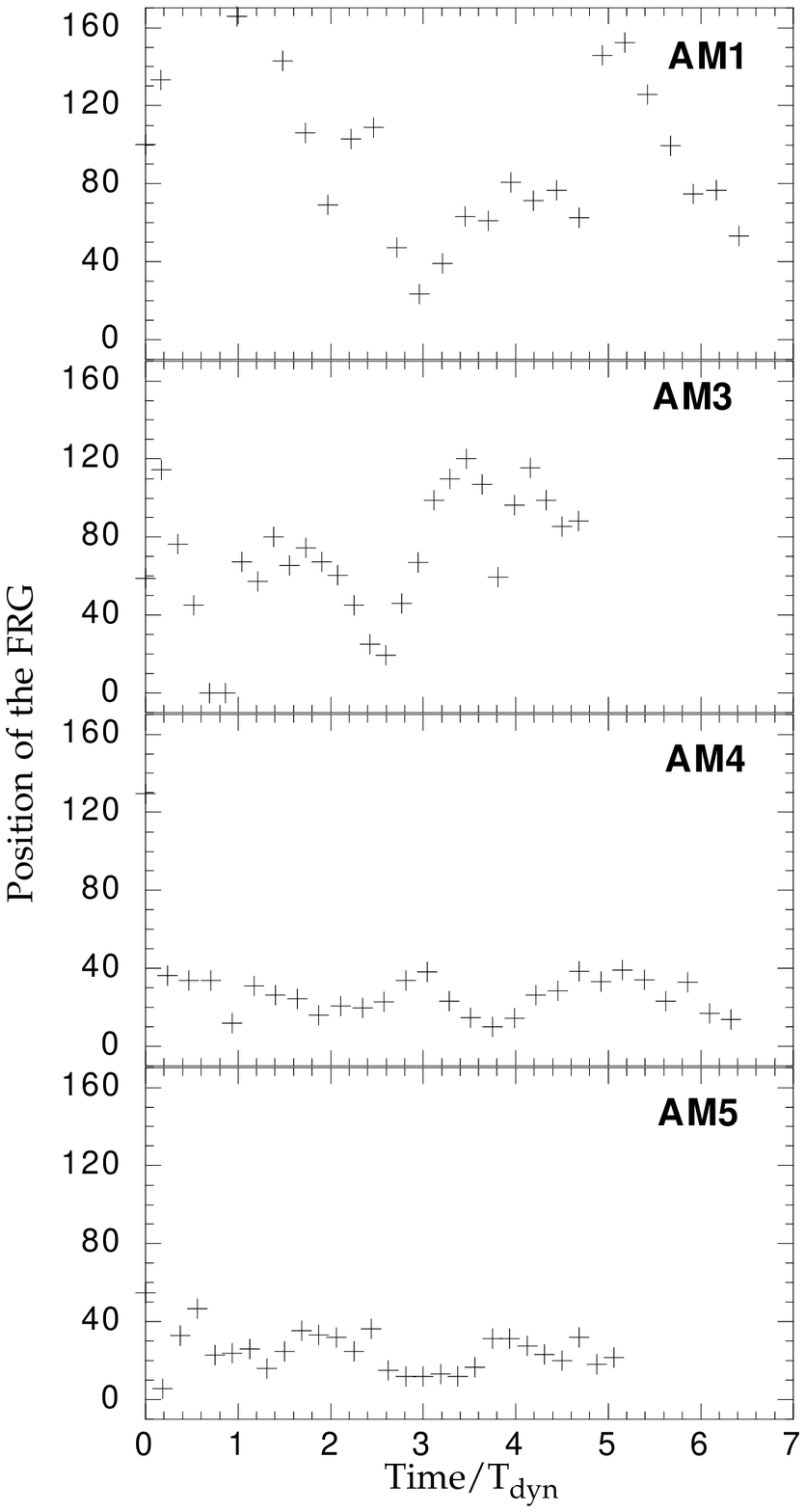,width=8.7cm,clip=true}
\figure{7.a}{Position of the FRG as a function of time for four isolated
cluster simulations. For comparison purpose, the time is scaled to the
dynamical time scale. The run ids. are on the top right of each panel
}
\endfig

\begfig 0.cm
\psfig{figure=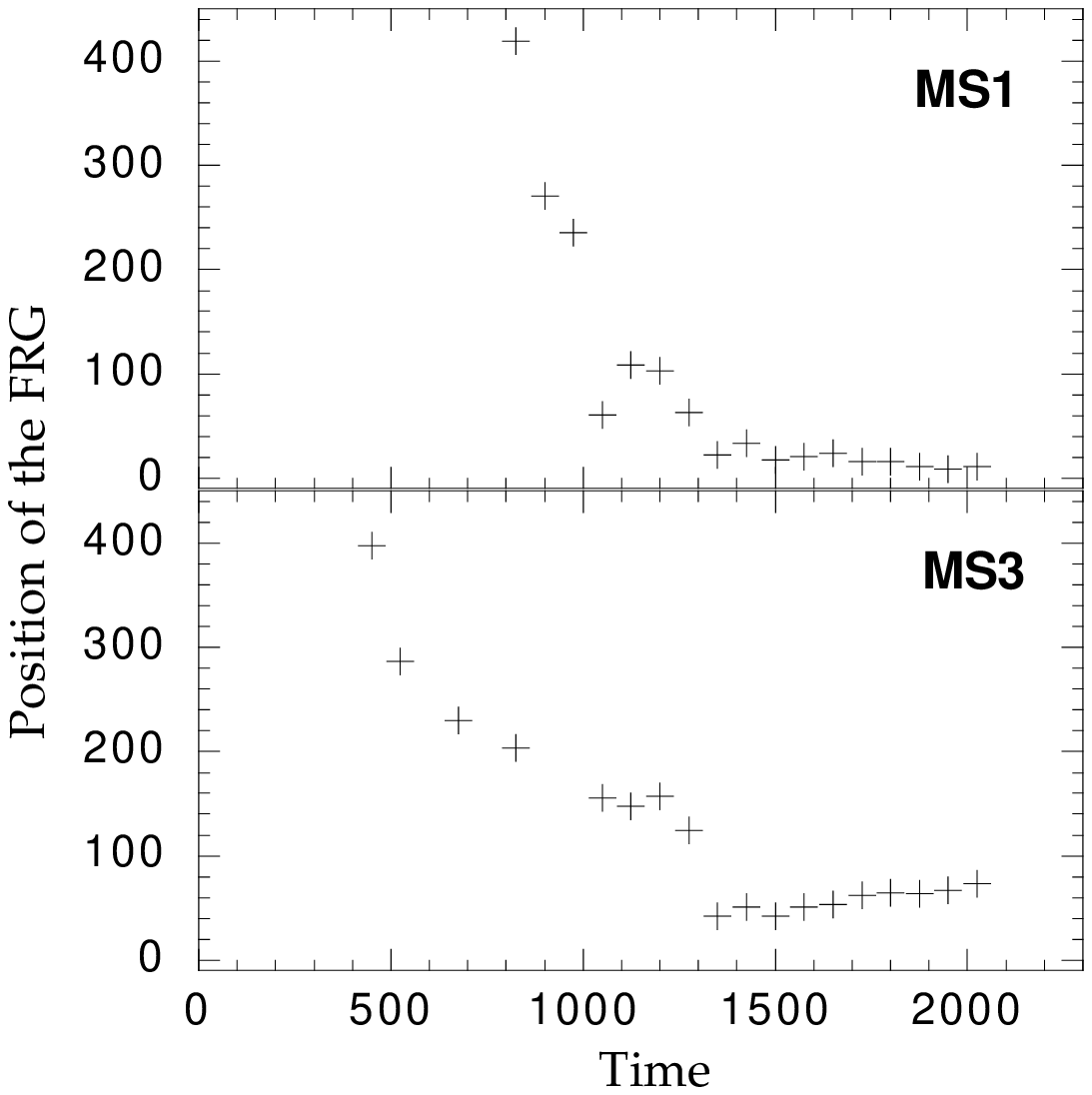,width=8.7cm,clip=true}
\figure{7.b}{Same as Fig. 7.a but for the simulations of colliding clusters.
The time scale is in our simulations units
}
\endfig

\begfig 0.cm
\psfig{figure=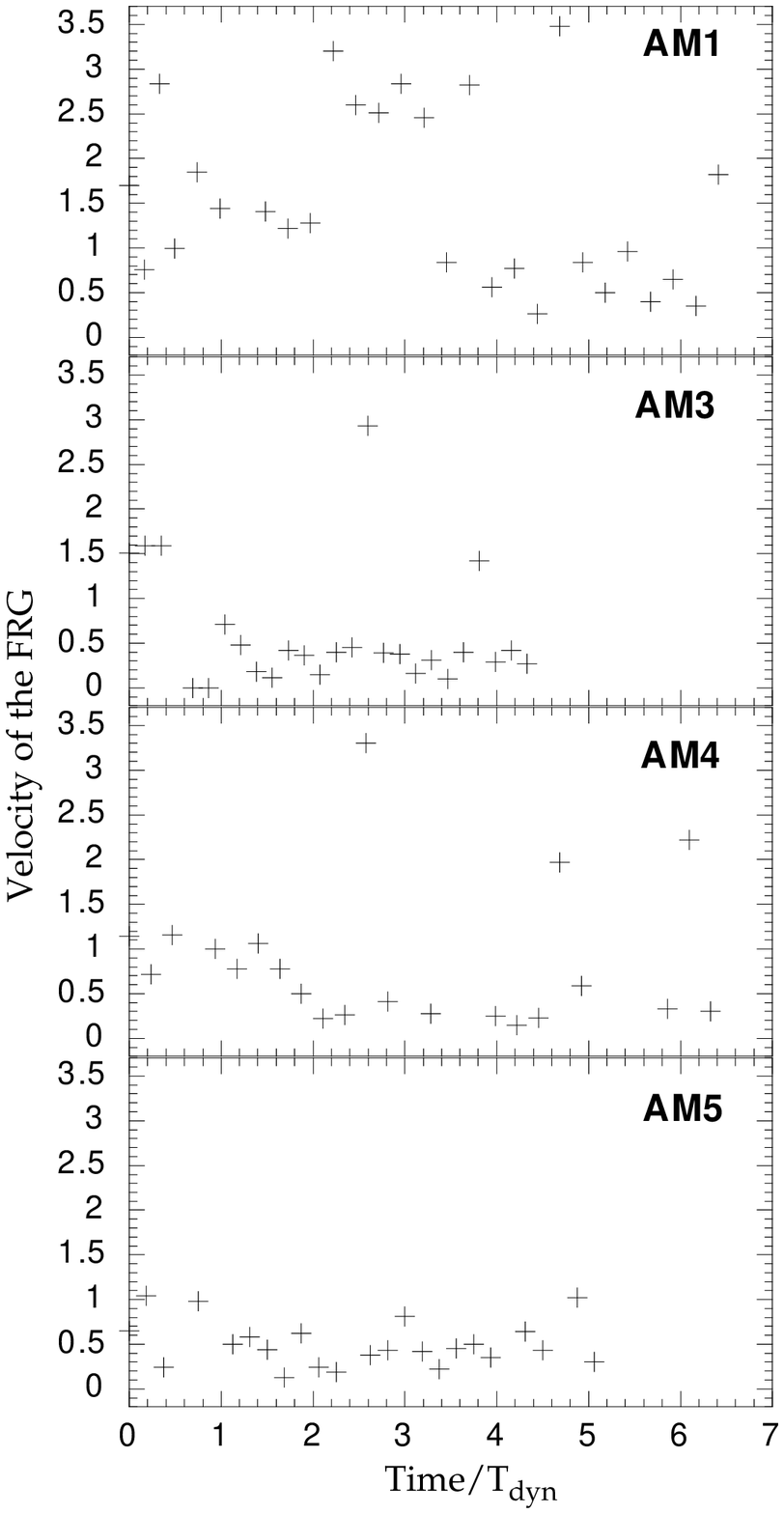,width=8.7cm,clip=true}
\figure{8.a}{Velocity of the FRG as a function of time for four isolated
cluster simulations. For comparison purpose, the time is scaled to the
dynamical time scale
}
\endfig

\begfig 0.cm
\psfig{figure=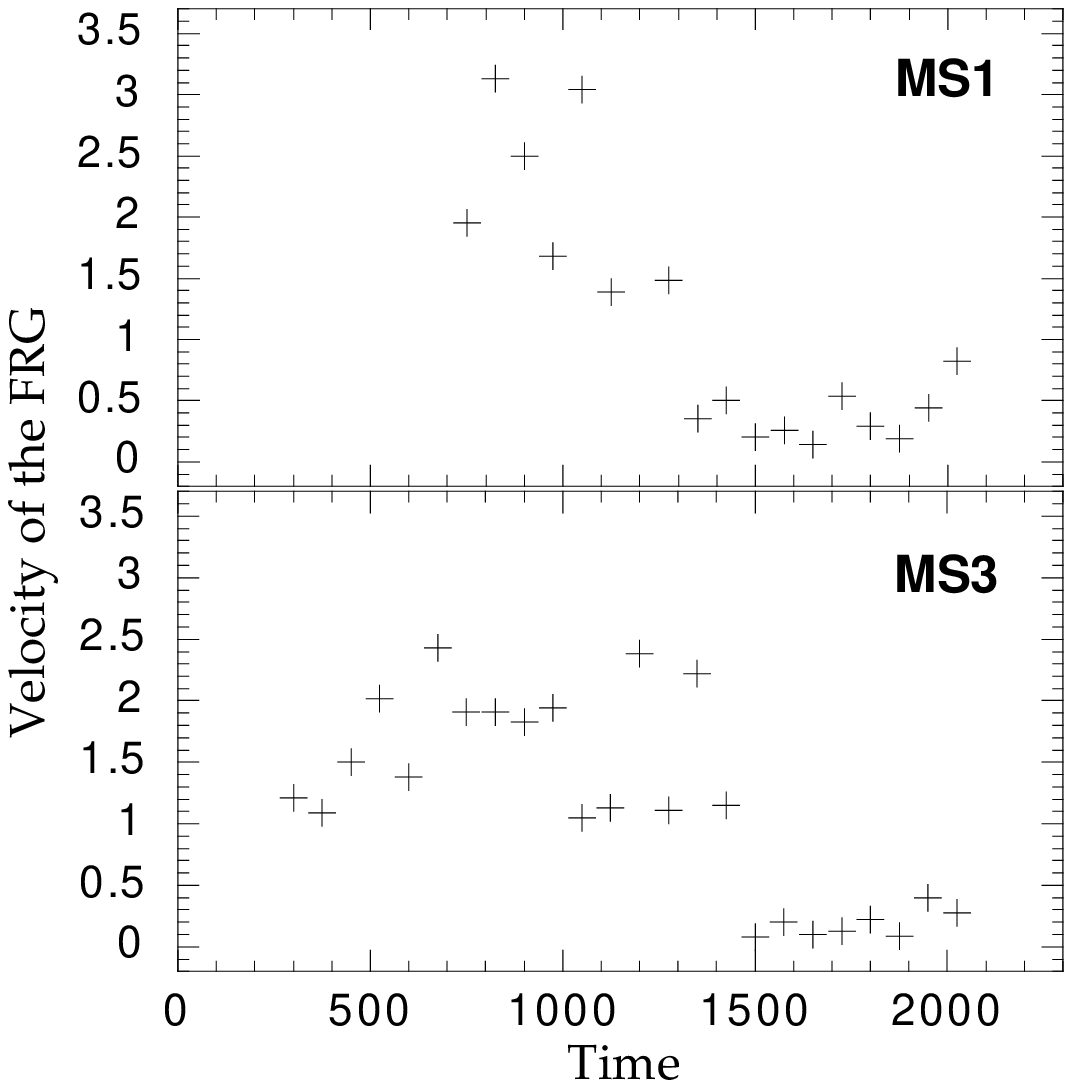,width=8.7cm,clip=true}
\figure{8.b}{Same as Fig. 8.a but for the simulations of colliding clusters.
The time scale is in our simulations units
}
\endfig

The FRG is not always formed already
in the centre. In the simulations AM1 and AM3, the FRGs are formed at
more than $\sim$ 200 kpc from the centre of the cluster. In the simulations
AM4 and AM5 the FRGs are formed in the core, at about 80 kpc from the centre.
The FRGs of simulations AM1 and AM3 fall towards the centre and remain in the
core.

In all simulations, the FRGs oscillate around the centre of the cluster.
However, the amplitude of this oscillation depends strongly on the
amount of ICM. In the simulation AM1, initially without ICM, the amplitude
of the oscillation reaches 160 kpc even after 5 $t_{\rm dyn}$ (or
9.6$\times10^9$yr). Increasing
the ICM, results in a decreasing oscillation amplitude; the simulation
AM3 reaches at most 120 kpc, and the simulations AM4 and AM5, with
the highest ICM, have maximum amplitude of $\sim$ 40 kpc.

Likewise, the FRGs velocity decreases with increasing ICM, although in
a less dramatic way (Fig 8.a). The FRG velocity in simulation AM1 goes from
about 150 to 350 km/s and finally remaining at 100 km/s. In the simulations
AM3, AM4 and AM5 the FRGs have an initial velocity of about 150--200 km/s
that falls to 80 km/s after 2 $t_{\rm dyn}$.

For the simulations MS1 and MS3 we follow one of the FRGs (there is
one for each subcluster). The position and velocity are relative
to the centre of mass of the whole system (Figs 7.b and 8.b). In both
simulations the FRGs fall to the centre in $\sim 7\times10^9$yr
where they merger with the other subcluster FRG. After merging
their behaviour is similar to the FRGs from the isolated cluster
simulations. The FRGs of simulations MS1 and MS3
remain at a distance of 30 and 70 kpc from the centre respectively,
with a peculiar velocity of $\sim$ 70 km/s.

It is interesting to note that when the FRGs of the colliding
subclusters are merging they have a higher peculiar velocity. In the
simulation MS1 (head-on collision), at 1300 time units the FRGs
are almost merged and have a peculiar velocity of 200 km/s. In
simulation MS3 (off-centre collision) we observe the same, the FRGs
have a velocity of 160 km/s relative to the centre of the cluster
just before finishing merging.

The above values are three dimensional. When projecting the cluster
on the plan of the sky and taking only the line-of-sight velocity,
the FRG will usually appear to be closer to the centre and with a
lower peculiar velocity.
These results show that the ICM is efficient to produce the dynamical
friction on the massive galaxies. Notice however that the position
is more strongly affected than the velocity of the galaxies.

\titleb{The mass function}

We have fitted the differential mass function using the Schechter luminosity
function at every 75 time-units in order to determine its evolution. Due
to the small number of objects in our simulations, the fits were done
by maximization of the Likelihood using a Poisson distribution for
the probability of having $n$ galaxies in a given interval $M, M+{\rm d}M$
where the expected number of galaxies is given by the Schechter distribution.
The errors on the fitting parameters are estimated with a Bootstrap (or
re-sampling) technique.

The most notable fact observed in our simulations is that there is {\it no}
or very little evolution
of the mass function of a cluster. This result applies for isolated
clusters as well as for the simulations of merging substructures. There are, 
however,
three factors that must be taken into account before analysing
this result. First, the mass range in the simulations is only a factor 12.5
and, second, the number of galaxies varies between 60 to
$\sim$ 25 from the beginning to the end of a simulation. Third,
as explained in \S 2.3, what we identify as a `galaxy' is 
actually its central part. More details of the evolution of the
mass function are given elsewhere (Lima Neto, 1996).

\titleb{Substructures}

\begfigwid 0.0cm
\psfig{figure=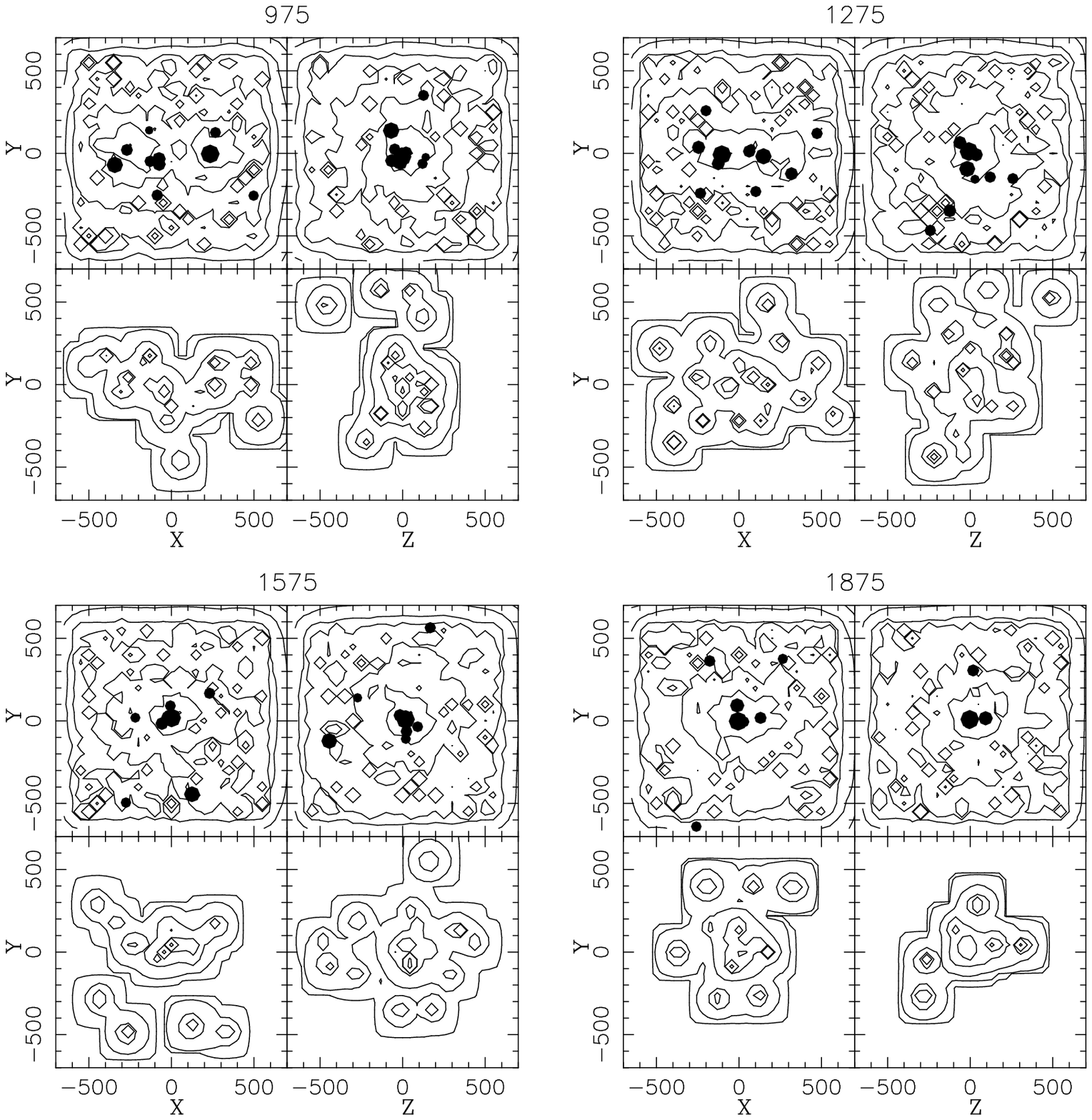,width=18cm}
\figure{9.a}{
Projected isodensities of the ICM compared to the distribution 
of galaxies of simulation MS1.
The upper two panels are two orthogonal projections of the ICM, below
are the corresponding isopleths of galaxy counts. The dots superposed to
the ICM isocontours represent the most massive galaxies.
Above each picture is the time.
}
\endfig

We have compared the distribution of the galaxy counts with the distribution
of the diffuse invisible matter (Figs. 9.a and b). The distribution of the
ICM (here, the dark
collisionless particles) should be similar to an hypothetical
emissivity map of X-rays. This is so supposing that the X-ray
emitting plasma traces the cluster gravitational potential. Such
hypothesis is justified due to the very short relaxation time
scale of the X-ray emitting plasma compared to the dynamical time scale
of the cluster itself.

In the simulations of isolated clusters with a dark ICM (AM3, AM4 and AM5), the
projected isodensity curves are relatively spherically symmetric although
rather noisy (Fig. 9.b). Small secondary maxima can be seen, always related to a
concentration of galaxies. Isopleths of galaxy counts show more structures
than the ICM on all simulations, as it seems to be the same case with
real clusters (e.g. Baier 1983). Globally, all simulated clusters show
some degree of subclustering.
This is of course higher when we have two substructures colliding but the
subclustering is nevertheless visible even in the simulations of isolated
clusters.

An interesting fact is that in some projections of the clusters the FRG
does not coincide with the maximum of the ICM density. For instance,
in one case we have the second more massive galaxy at about 50~kpc from centre
of the projected ICM distribution and the FRG at about 200~kpc. This happens
at a
time (in physical units) of 11.8$\times 10^9$ years, counting from the
beginning of the simulation. Looking at the evolution of this particular
simulation, we notice that what happens is that the galaxy closest to
the centre was the FRG while the second ranked galaxy was farther away.
However, the later merges with another large galaxy at $T\approx
11.0\times 10^9$
years, and becomes the FRG. Therefore, the offset between the centre of the
ICM and the FRG is, in this case, a transient effect which lasts a little
less than $10^9$ years, the time it took for the FRG fall to the centre
of the cluster. 

\titleb{Rotation in clusters of galaxies}

\begfigwid 0.0cm
\psfig{figure=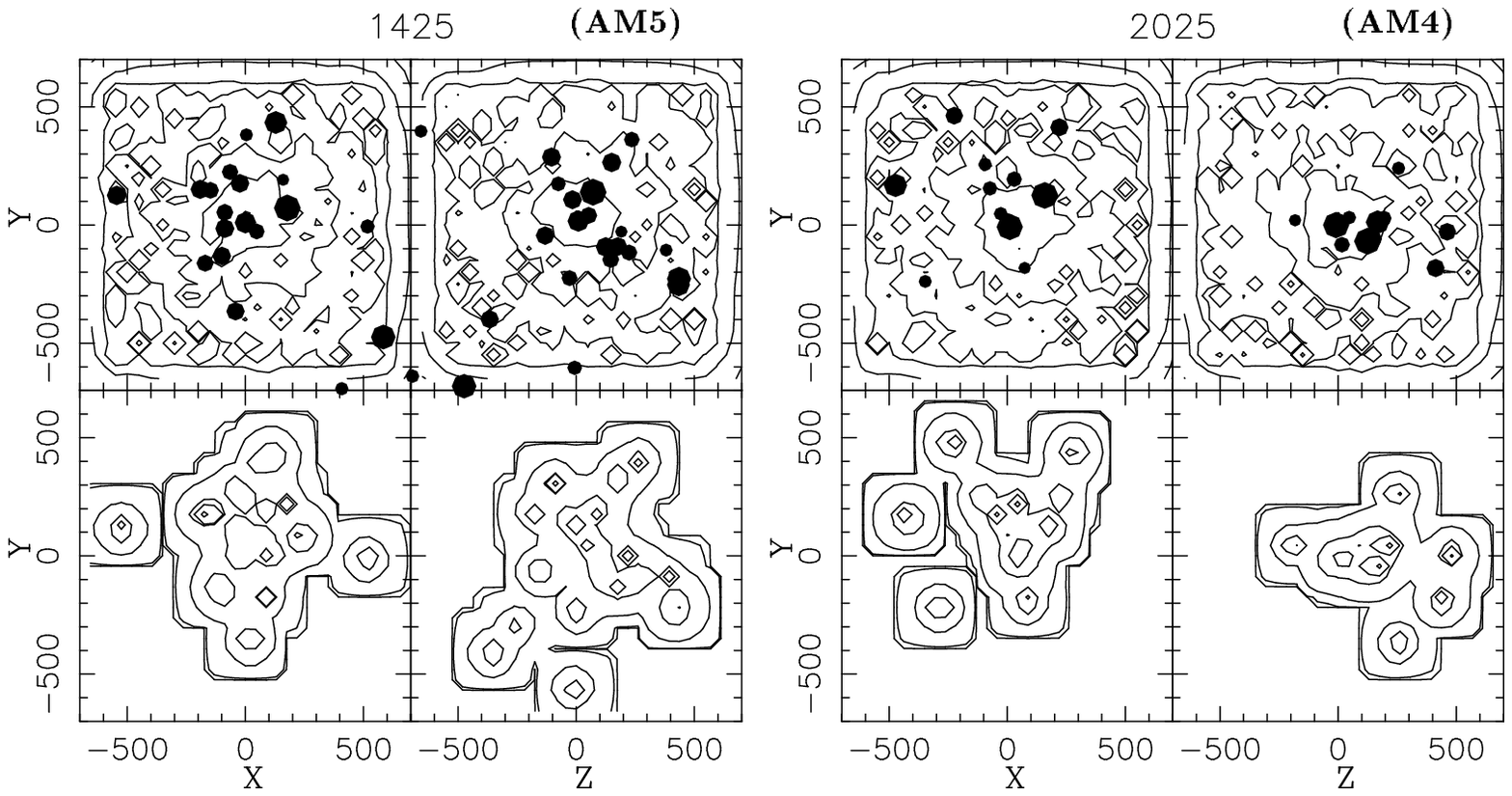,width=18cm}
\figure{9.b}{Same as Fig. 9.a but for simulations AM5 and AM4.
}
\endfig

When two gravitationally bound subclusters collide with a non zero impact 
parameter, the whole system may have a significant amount of angular
momentum. The question is if a global rotation of the clusters can be
detected. Figure 10 shows the plot of the galaxy line-of-sight velocity as a
function of the position inside the cluster for the simulation MS3.

Given optimal conditions, i.e. the observer being located near the orbital 
plane of the subclusters and observing them as their projected separation
is about the size of their diameter (when the subclusters are just 
`touching' each other),
one can clearly detect the rotation of the whole cluster around its geometric 
centre. However, after the two subclusters merge completely, the rotation is 
less detectable but still present.
The angular momentum of the galaxies relative to the 
centre of the cluster is transferred to the massive ICM.
That is similar to what happens with two spiral galaxies of about the same 
mass that merge and form an elliptical galaxy. The angular momentum of the 
discs and the orbital angular momentum are transported to the massive halo
(Barnes, 1988).

We computed the dimensionless spin-parameter defined by
$$ \lambda = G^{-1} J (|E| M^{-5})^{1/2}\, , \eqno\autnum$$
where $J$, $E$ and $M$ are the total angular momentum, energy and mass,
respectively. The isolated and the head-on collision simulations have
$\lambda \approx 0.02$--0.06. It is not zero due to the random fluctuation
when generating the initial conditions that produces some trifling rotation.
On the other hand, the simulations of subcluster collision that are
initially in an elliptical orbit,
specially MS3 and MS4, have $\lambda \approx 0.3$, while the
simulation MS2 has $\lambda \approx 0.5$. These values are high in comparison
to the predicted rotation of clusters produced only by tidal torques
which is about $\la 0.2$ (Efstathiou \& Barnes 1984). As a reference,
a spiral galaxy has $\lambda \approx 1$.

\titlea{Discussion}

A number of $N$-body simulations of poor clusters of galaxies using particle 
methods, has been performed by diverse authors
recently. Our models relate closely to the simulations of Barnes (1989),
Malumuth (1992), Funato et al. (1993), and Bode et al. (1993, 1994).

Similarly to Funato et al. (1993), the stripping of particles that were
initially bound to the galaxies is very important. In our simulations,
more than half of the particles initially in the galaxies are, at the end
of the run (or $\sim 12 \times 10^9$ yr), on a huge
envelope around the central dominant galaxy. That means that starting with
an ICM ($M_{\rm ICM}/M_{\rm tot}$) of 67\% we arrive at an ICM of 83\% by
the end of the simulation.

Contrary to the results of Bode et al. (1994), the number of mergers in
our simulations increases with increasing initial ICM. That probably comes
from the way the clusters are constructed. When Bode et al. increase their
ICM (noted $\beta$ in their paper) they actually reduce the mass in the 
galaxies
and, consequently, the radius thus reducing the merging cross section of the
galaxies. On the other hand, in the simulations here we keep the mass and
radius of the galaxies, augmenting the ICM mass and hence the velocity
dispersion. The overall effect is to increase the dynamical friction which
facilitates the formation of the central dominant galaxy by cannibalism.
The merging rate
is thus higher because mergers occur mainly with the central cannibal galaxy.

The collision of subclusters has been intensely investigated with the help of
numerical 
simulations during the last few years (e.g. Roettiger et al. 1993, Jing et al. 
1995, and Nakamura et al. 1995). The simulations of Roettiger et al. (1993) 
are based on
a hybrid hydrodynamics--tree-code appropriate to follow the X-ray emitting gas
but lacking enough particles to resolve the galaxies. Jing et al. (1995) 
P${}^3$M simulations have pertinent cosmological initial conditions but again 
do not resolve the individual galaxies.

\begfigwid 0.0 cm
\psfig{figure=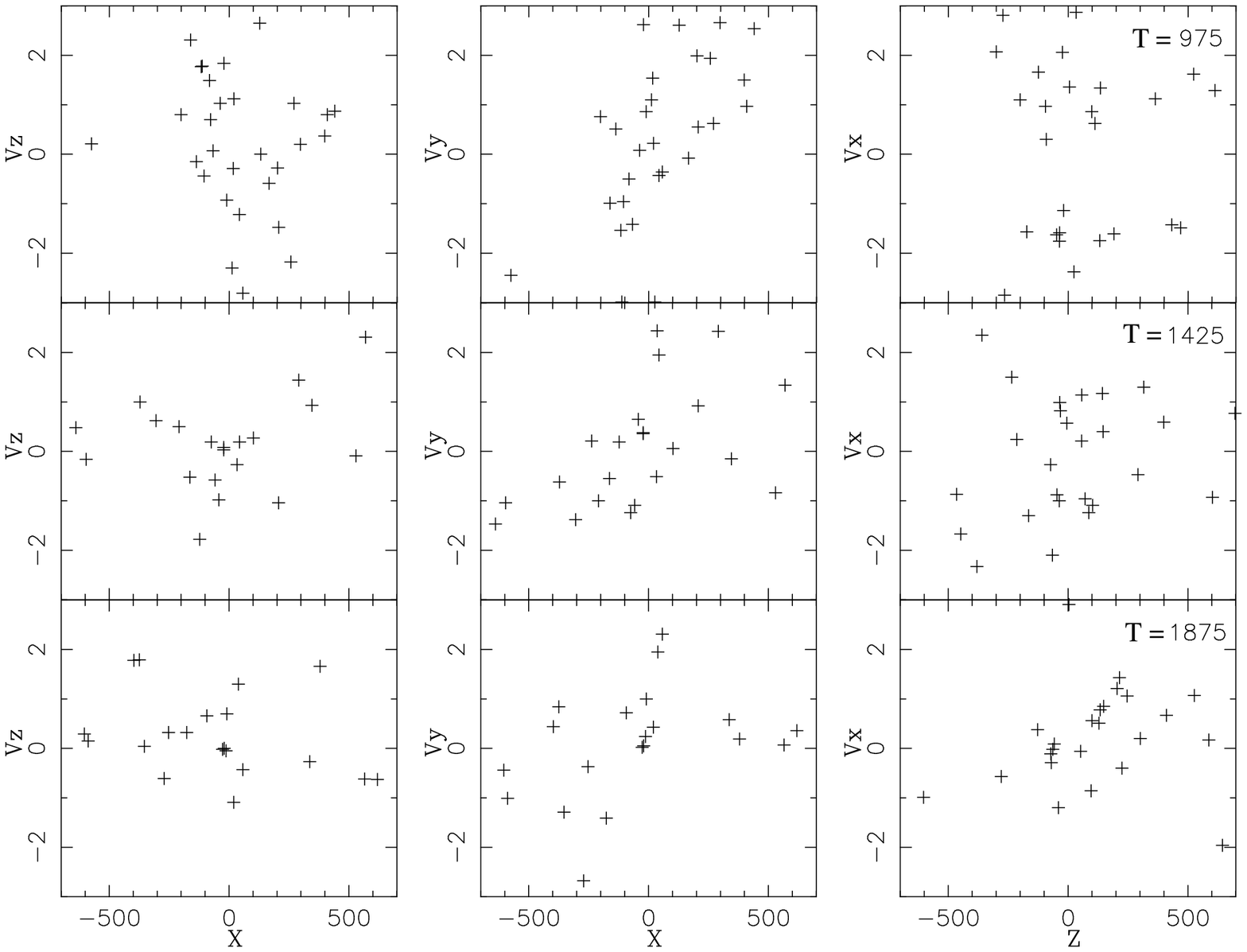,width=18cm}
\figure{10}{
Line-of-sight velocity as a function of the projected distance from the centre 
of the cluster for the simulation MS3. Each column represents a perpendicular
projection of the cluster. The time is shown on the panels on the right column.
A rotational pattern can be seen when the 
cluster is observed from near the orbital plane ($Y$--$Z$)}
\endfig

Nakamura et al. (1995) use direct summation to solve the equations of motion
and, also, do not resolve the 
galaxies. They conclude from their experiments that the initial distribution of 
the invisible matter is relevant to determine the time scale for erasing the 
substructure feature after the collision of the substructures. They estimate 
that this can take more than
$4\times10^9$yr after the first encounter. In our simulation MS1 (head-on 
collision of subclusters) it takes about $2.6\times10^9$yr after the first 
encounter for the substructures of the mass distribution to be erased. This 
difference seems to be simply due to the slightly lower relative velocity 
between the subclusters that we adopted, 1.5 instead of 1.92 that we would have 
using their prescription (based on having the clusters at rest at infinity). In 
other words, our simulation has a higher binding energy and the double peak is 
washed out faster.

Malumuth (1992) simulated clusters of 100 galaxies with 50\% of the mass
in the ICM using an `explicit physics' code (i.e. dynamical friction,
merging, and tidal stripping are treated in a statistical way) in order to
study the peculiar velocities of cD galaxies. The position and velocity
of the FRGs in our simulations are in agreement with his work, where
78\% of his cD galaxies lie closer than 100 kpc from the centre and 72\%
have a velocity lower the 150 km/s. Malumuth also suggested that cD
galaxies with high peculiar velocity should be found preferentially
in clusters with substructures. Indeed, in our simulations the FRGs with
higher peculiar velocity are found in colliding clusters (where substructures
are most significant) at the moment that the subclusters are merging.
However, after the FRGs from both subclusters merge and form a single
giant galaxy, it quickly fall to the centre and is braked there by
dynamical friction. Thus, a cD galaxy could be detected with a high peculiar
velocity during an interval of $\sim$ 1--3$\times 10^9$yr, taking into account the
time it takes for the merging and braking by dynamical friction.

Although the distribution of galaxies is clumpier than the ICM, the galaxy
distribution reflects the total matter distribution. Thus, if the
intra-cluster plasma traces the gravitational potential, the galaxy distribution
should trace the X-ray emission. This seems to be indeed the case in real
poor clusters (Dell'Antonio et al 1995).

The collision and merger of two subclusters initially in a highly elliptical
orbit produces a final cluster with an amount of rotation detectable
with the line-of-sight velocity of the galaxies. Although the observer
must be close to the orbital plane of the subclusters, the analysis
of clusters classified as elongated, flattened or bi-modal may
show a rotational pattern of the galaxies around the centre of the cluster.
Since tidal torques are not enough to produce the rotation
on this scale (Efstathiou \& Barnes, 1984), the observation of rotating
clusters would
support hierarchical cosmological theories where smaller units merge to
form bigger ones. At least one cluster, SC 0316-44, has a velocity pattern
that corresponds closely to our simulation MS3 (Materne \& Hopp 1983).
It is an elongated cluster containing two massive galaxies,
a cD and a D. This may be
a case of two subclusters spiraling towards each other.

It is interesting to note that the collision of substructures does
not seem to affect the rate that galaxies merge. As stated above,
mergers occur preferably with the central cannibal galaxy rather than
pairwise (as already noted by Bode et al. 1994). Since the cannibal
galaxy formation is faster than the coalescence of substructures, it
is natural to understand the small (or none) effect of subcluster
collisions on the merging rate of galaxies. Thus, each substructure
behaves as an isolated mini cluster, forming its own FRG and
cannibalizing the smaller ones almost independently of its surroundings.

At last, increasing the number of particles from 46\,000 to 70\,000 shows no
effect on the evolution of our simulations.

\titlea{Conclusion}

We have performed $N$-body simulations of poor clusters using enough
particles to model the internal structure of the galaxies in a
self-consistent way.
The main results of this work are resumed as follows.

The merging of substructures does {\it not\/} increases the
merging rate of galaxies in clusters of galaxies. However, it seems to be an
efficient mechanism to produce cD galaxies near the centre of the
potential wells, where the ICM and the cD galaxy of each substructure have merged.
Moreover, the growth rate of the FRG is strongly correlated with
the merging rate in a cluster.

The position of the first ranked galaxy does {\it not\/}
coincide always with the maximum density of the dark matter, even
in an isolated cluster. Assuming
that the intra-cluster X-ray emitting gas traces the gravitational
potential of a cluster (dominated by the invisible matter) this
implies that the position of the FRG may not match the maximum X-ray
emission of a cluster.

The merger of substructures (or merger of poor clusters)
may produce situations where the position of the FRG does not
coincide with the position of the deepest point of the potential
well produced by the dark matter. This offset is most visible just
after the merging of the dominant galaxies of each substructure.
This situation can also be observed in isolated clusters, when
the dominant galaxy is still oscillating around the centre of the
cluster.

Our simulations suggest that the initial mass function of
the galaxies is narrowly related to the formation and development
of an important central FRG with cD characteristics. A steep mass
distribution function facilitates the creation of a central dominant
galaxy whereas a flat distribution tends to slow its formation.

The shape of the mass distribution function shows no or only
a very small evolution during the life span of a cluster, even
when there are pronounced substructures. This result, however,
should be regarded carefully since the mass range in these
simulations was small ($\sim$ 12) compared to a real cluster of
galaxies.

Finally, the merger of two subclusters with a non zero impact parameter
can produce a detectable (with
the observation of the line-of-sight velocity), rotating
cluster of galaxies.

\acknow{One of us (G.B.L.N.) thanks the support from the  Alexander von Humboldt
Foundation. The computations were performed on the Crays J90 and EL92
of the {\it Astrophysikalisches Institut Potsdam}.
}

\begref{Bibliography}

\ref Albert, C.E., White R.A., Morgan W.W., 1977, ApJ 211, 309
\ref Baier F.W., 1983, Astron. Nachr. 304, 211
\ref Baier F.W., Lima Neto, G.B., Wipper H., Braum M., 1996, Astron. Nachr.
317, 77
\ref Bahcall N.A., 1980, ApJ Lett. 238, L117
\ref Barnes J., 1988, ApJ 331, 699
\ref Barnes J., 1989, Nature 338, 123
\ref Barnes J., Hut P., 1986, Nature 324, 446
\ref Beers T.C., Forman W., Huchra J.P., Jones C., Gebhardt K., 1991,
Astron. J. 102, 1581
\ref Bode P.W., Cohn H., Lugger P., 1993, ApJ 416, 17
\ref Bode P.W., Berrington R.C., Cohn H.N., Lugger P.M., 1994, ApJ
433, 479 
\ref Chincarini G., Vettolani G., de Souza R.E., 1988, A\&A 193, 47
\ref Dell'Antonio I.P., geller M.J., Fabricant D.G., 1995, AJ 110, 502
\ref Efstathiou G., Barnes J., 1984, in: Formation and evolution
of galaxies and large structures in the Universe, p. 361,
Audouze J. and Tran Thanh Van J. eds., D. Reidel Publishing Company
\ref Fabian A.C., Nulsen P.E.J., Canizares C.R., 1984, Nature 310, 30
\ref Fort B., Mellier Y., 1994, A\&AR 5, 239
\ref Funato Y., Makino J., Ebisuzaki T., 1993, PASP 45, 289
\ref Geller M.J., Beers U.G., 1982, PASP 94, 421
\ref Hernquist L., 1988, Comp. Phys. Comm. 48, 107
\ref Lima Neto G.B., 1996, Astr. Letts. Comm., submited
\ref Jing Y.P., Mo H.J., B\"orner G., Fang L.Z., 1995, MNRAS 276, 417
\ref Jones C., Forman W., 1990, in: Clusters of Galaxies, p.~257,
Oegerle W.R., Fitchett M.J. and Danly L. editors, Cambridge Univ. Press
\ref Malumuth E. M., 1992, ApJ 386, 420
\ref Materne J., Hopp U., 1983, A\&A Lett. 124, L13
\ref McGlynn T.A., Fabian A.C., 1984, MNRAS 208, 709
\ref Merritt D., 1983, ApJ 264, 24
\ref Morgan W.W., Kayser S., White R.A., 1975, ApJ 199, 545
\ref Nakamura F.E., Hattori M., Mineshige S., 1995, A\&A 302, 649
\ref Ostriker J.P., Tremaine  S.D., 1975, ApJ Lett. 202, L113
\ref Roettiger K., Burns J., Loken C., 1993, ApJ Lett. 407, L53
\ref Sarazin C.L., 1988, ``X-ray emissions from clusters of galaxies'',
Cambridge University Press
\ref Schechter P., 1976, ApJ 203, 297
\ref Ulmer M.P., Wirth G.D., Kowalski M.P., 1992, ApJ 397, 430
\ref West M.J., Oemler A., Dekel A., 1988, ApJ 327, 1
\ref Zabluboff A.I., Zaritsky D., 1995, ApJ Lett. 447, L21
\endref
%
\bye